\documentstyle[12pt,epsfig]{article}
\begin{document}

\vbox{\vskip 2truecm}

\centerline{\ \ \ \ \ \ 
\ \ \ \ \ \ \ \ \ \ \ \ \ \ \ \ \ \  
\ \ \ \ \ \ \ \ \ \ \ \ \ \ \ \ \ \ 
\ \ \ \ \ \ \ \ \ \ \ \ \ \ \ \ \ \  Crete 96-12}

\bigskip

\bigskip

\bigskip

\centerline{The Standard Model process 
$e^+e^-\rightarrow \nu\bar{\nu}b\bar{b} $ and its Higgs signal at LEPII}

\bigskip

\centerline{D. Apostolakis, P. Ditsas}
\centerline{Department of Physics, University of Crete and}
\centerline{Foundation of Research and Technology (FORTH)} 
\centerline{Heraklion, Crete, Greece}
\centerline{S. Katsanevas}
\centerline{Department of Physics, University of Athens}

\vbox{\vskip 5truecm}

\vfill

\centerline{\bf Abstract}
\bigskip
We present and study the results for the standard model process 
$e^+e^-\rightarrow \nu\bar{\nu}b\bar{b}$\ 
at c.m energies 150$\leq \sqrt{s} (GeV) \leq $ 240 and
for Higgs boson masses $80 GeV \leq m_H \leq 120 GeV$, obtained from all 
tree-level diagrams and including the most important radiative corrections.
The matrix elements have been calculated by the 'spinor bracket' method
without neglecting masses, while the phase space integrals by an 
importance sampling  Monte Carlo numerical integrator.
The $\sqrt{s}$ dependence and the interference properties of the Higgs boson 
contribution and of various coherent background contributions to the 
total cross section are examined. The effect of the QED initial state
radiative corrections is estimated.  The important differential distributions 
for the Higgs boson and the background components are studied, providing 
information usefull for choosing cuts in Higgs searches. We also examine 
the effect of a minimal set of cuts and evaluate the importance of the 
WW fusion for detecting a higher mass Higgs boson at LEPII.
\vfill

\bigskip
\noindent
\section{Introduction}
\bigskip

Increasingly precise LEP I data \cite{1}, combined with an evidence for the 
top quark of mass $176\pm 15$ GeV \cite{2}, have made feasible a very important
$-$and impressively successful$-$ test of the Standard Model (SM) to an
accuracy level of one-loop quantum corrections \cite{4}, \cite{5}. These data
are also compatible with the Minimal Supersymmetric extension of the SM
(MSSM \cite{5}, which (at a price of introducting many more particles) provides
perturbative stability against radiative corrections in a grand 
unification scheme.

Yet, comparison to presently available data and requirements of theoretical
consistency do not restrict sharply the mass of the SM Higgs boson,
and even more so the parameters of the Higgs sector of the MSSM \cite{6}.
On the experimental side, LEP I
collaborations converge slowly towards a combined lower experimental bound 
$m_H \geq  65$ GeV for the mass of the SM Higgs boson \cite{1}: given
this bound, the other results of LEP I are well-known to be rather
insensitive on $m_H$. As for the theoretical constraints,
the radiatively corrected SM Higgs potential $V[\phi]$ develops an 
instability \cite{6a}, 
and depending on the assumptions about the cure of this 
quantum catastrophe, some limits on $m_H$ are implied. E.g. assuming that
the
instability is avoided perturbatively by onset of new physics at
$\Lambda\approx 10^3$ GeV (respectively $\Lambda\approx 10^4$ GeV)
implies, for $m_t=174$ GeV and $\alpha_s(m_z)=0.118$, a lower
limit $m_H>75$ GeV (respectively $m_H>100$ GeV) \cite{7}. A change 
$\Delta m_t$ of the top quark mass induces a change 
$\Delta m_H\approx 2\Delta m_t$ of such a limit \cite{7}.
 On the other hand, assuming
the MSSM with $m_t\approx 170$ GeV and supersymmetry breaking scale 
$m_s\approx 10^3$ GeV implies an upper limit $m_H<125$ GeV on the 
lightest mass Higgs boson, the limit changing with $m_t$ according to 
$\Delta m_H\approx\Delta m_t$ \cite{8}. The same upper bound (for 
$m_t\approx 170$ GeV) was found from a low energy renormalization group
analysis of the Higgs sector of the `next-to-minimal' supersymmetric
SM \cite{9}. Actually, because of the proximity of a fixed point in the low 
energy evolution of $m_t$, one would expect that $m_H$ is well below this 
upper bound, expecting e.g. $m_H \leq  105$ GeV for
 the above values of $m_t$ and
$m_s$ \cite{10}.

Detection of at least the lightest Higgs boson (HB) and experimental study of 
its properties is presently among the highest priority goals in particle
physics. Yet,  LEP II will provide the only opportunity in the near future to 
search directly for a HB of mass $m_H>65$ GeV. It is thus very important to
have complete and carefully analysed results for all $e^+e^-$-collision
SM and MSSM processes with an `uncoverable' HB signal 
at 170-210 GeV c.m. energies. For the first time many of these processes
will have 4 fermion
 final states, arising partly through 2 intermediate 
 heavy bosons, and we should learn to optimize the search for the HB
in such processes and to push the explorable $m_H$ range as high as possible.
A precise knowledge of the  experimental background to these processes
is, of course, also much needed.

In this work we assume the validity of the SM at LEP II energies.
Since the SM Higgs boson detectable at LEP II decays predominantly to 
$b\overline{b}$, the most promising SM processes are

\begin{eqnarray}
e^+e^-\to \ell^+\ell^- b\overline{b},\,\,\, \nu\overline{\nu}
b\overline{b},\,\,\, q\overline{q}b\overline{b} \nonumber
\end{eqnarray}
where $\ell = e$ or $\mu$, summation over the neutrino species is 
understood, and $q\overline{q}$ is a quark-antiquark pair generating jets
just like the $b\overline{b}$ pair does.
In the present study we concentrate on the 
 complete tree level calculation, including 
approximatively the most important radiative corrections, of the process
\begin{eqnarray}
e^+e^-\to \nu\overline{\nu}b\overline{b} 
\end{eqnarray}
and examine some problems of extracting the HB signal from
it at LEP II despite a large background.
We should mention that process (1) is of interest at LEP II
for another reason: it will constitute an important background in the 
search for the process $e^+e^-\to \chi\chi b\overline{b}$, where $\chi$
would be the lightest neutralino, stable when R-Parity is respected. 
Both $\nu$'s and $\chi$'s (if they exist) will be `felt' only 
through energy-momentum conservation.

An intermediate SM Higgs boson contributes to process (1) through the 
so-called Higgsstrahlung and $WW$ fusion diagrams of Fig. 1. This HB signal
is embedded in a  coherent electroweak background arising from the 20
diagrams fo Fig. 2. These can be classified as
\begin{itemize}
\item the $e$-exchange diagrams of Fig. 2a, where $ZZ$ or $Z\gamma$
coexist and cascade on the $s$-channel
\item the $WW$ multiperipheral and fusion diagrams of Fig. 2b
\item the $W$- and $e$-exchange diagrams of Fig. 2c, with a radiation
of $\gamma$ or $Z$
\item the $s$-channel $e^+e^-$ annihilation diagrams of Fig. 2d, with 
$Z$ radiation from a final fermion.
\end{itemize}

For reasons of brevity, the background processes of Fig. 2a, b, c, d will be
called respectively $ZZ/\gamma$ cascade, $WW$ fusion, $W$ exchange 
and $Z$ radiation processes. 
We work in the t' Hooft-Feynman gauge, but have ignored diagrams
with a coupling of the Higgs boson or of Goldstone bosons to
$e^{\pm}$, $\nu_e$, $\overline{\nu}_e$ leptons,
 since they contribute insignificantly.

Experimentally, one must extract the HB signal form  all final states 
consisting of a pair of  imperfectly  $b$-tagged jets + missing
products. 
The important components of the  incoherent background, denoted 
$B^{\gamma}_{QCD}$, $B_{WW}$, $B_W$, $B_{\ell\ell}$ and $B_{\mu\mu}$ for 
brevity, correspond to the following processes:
\begin{itemize}
\item $B^{\gamma}_{QCD}$: $s$-channel (virtual $\gamma/Z$ mediated)
 production of $b\overline{b}$ or
$c\overline{c}$ jets (there is the possibility of 
misidentification of $c$ jets 
in $b$-tagging), accompanied by soft undetected
hadrons or by beam radiated photons escaping detection.
\item $B_{WW}$: $WW$ production, with one $W$ decaying to cs jets
and the other decaying undetected through a chain 
$W\to \nu\tau\to \nu\overline{\nu}+$ soft charged particles.
\item $B_{W}$: $We\nu$ events, with $e$ escaping undetected and $W$
decaying into cs jets.
\item $B_{\ell\ell}$, $\ell =e$ or $\mu$: 
$\ell^+\ell^-b\overline{b}$, $\ell^+\ell^-c\overline{c}$ final states, with 
the leptons escaping in the forward and/or backward direction.
\end{itemize}

It is evident that very efficient criteria of event acceptance must 
be invoked in order to isolate (if possible) the Higgs component at a 
detectable level.

In order to put the present study in a proper context, we recall briefly
previous work on $e^+e^-\to\nu\overline{\nu}b\overline{b}$. Starting
from \cite{14}, there have been several ambitious simulations of this process by
experimental groups, aiming at a realistic estimate of the SM Higgs boson
signal to background ratio at LEP II (e.g. \cite{15},\cite{16}). 
Yet, they 
ignored the $WW$ fusion diagram of Fig. 1, 
assuming that it was important only at higher energies. But as shown by our 
results, WW fusion is already important at LEPII energies for two reasons:
(a) it increases the signal by about 10\% around its maximum and (b) 
it increases by large factors the weak signal  around
the kinematical threshold for HZ production.
They also ignored 
parts of the electroweak background and evaluated the rest approximately,
invoking only $2\to 2$ body processes followed by appropriate decays.
Theorists, on the other hand, were contended to study 
$e^+e^-\to\nu\overline{\nu}H$ without quantitative studies of the 
background. They usually included the $WW$ fusion 
contribution to this $2\to 3$ body process 
\cite{17},\cite{18},\cite{19}, but stressed its 
importance only for $\sqrt{s} \geq  500$ GeV 
\cite{20},\cite{21}. It was only 
recently \cite{22a},\cite{22b} 
when our present work was well under way, that the process
$e^+e^-\to\nu\overline{\nu}b\overline{b}$ was calculated including all signal
and background contributions, with emphasis on results for $\sqrt{s}$ 
above 200 GeV.

Despite some overlap between \cite{22a},\cite{22b} 
and the present study, we decided 
to complete and publish our results for several reasons. We shall not dwell
on a trivial one, the need to check independently calculations involving
22 diagrams and multidimensional phase space integrations, since we find 
good agreement when calculating same things with same parameters. More 
importantly, in view of the necessity of event selection criteria
for improving the signal/background ratio, we put emphasis on studying the
detailed characteristics of the signal and the 
coherent background, including some differential cross sections not 
considered in \cite{22a},\cite{22b}. This information can be 
beneficially combined   
with experience about $B^{\gamma}_{QCD}$, 
$B_{WW}$, $B_W$ and $B_{\ell\ell}$ to choose a set of experimental cuts
aiming to extract a measurable HB signal.
We also introduce several quantitatively
important technical improvements over previous work
(approximative inclusion of the
QED  initial state radiative (ISR) corrections
and of important $QCD$ radiative corrections to the adopted HB
width and the Higgs$-b\overline{b}$ coupling, approximative
employment of an Improved Born Approximation), 
and examine the effect of a set of minimal cuts.
These results allow estimates of the SM Higgs boson
detectability at LEP II more realistic than previously.
As a by-product, we reevaluate the importance of the $W-W$ fusion for 
detecting a higher mass HB at LEP II, already noticed in 
\cite{22a},\cite{22b}.
We should mention once more that, for reasons of experimental accessibility,
all cross section presented below are summed over neutrino species, 
(in \cite{22b} cross sections with electronic neutrinos are emphasized).

\section{II.\ Method of calculation and computation}
\bigskip
The process $e^+e^-\to\nu\overline{\nu}b\overline{b}$ interrelates six
external spin ${1\over 2}$ fermions. For fixed momenta of these fermions,
the initial-spin averaged and final-spin summed matrix element squared is 
given by
\begin{eqnarray}
{1\over 4}
\sum_{\lambda}\left|\sum_{F}iM^F_{\lambda}\right|^2
\end{eqnarray}
where $\lambda$ runs over a basis of spin states for external fermions and 
$F$ runs over all Feynman diagrams contributing to the process. Being 
interested in the Higgs signal, we must avoid approximating the $b$
quark as massless: so we decided not to use this approximation for the
electrons either. We then choose a convenient spinor basis in accordance with 
\cite{23},\cite{24} and calculate analytic expressions for all $iM^F_{\lambda}$:
the remaining operations in (2) are performed, for each point of phase
space, by a computer program. (The advantages of this procedure in comparison
to the traditional `trace technique' are discussed in \cite{23},\cite{24}). 
Before 
describing the structural form of $iM^F_{\lambda}$, we would like to 
briefly motivate (in a novel way) and present the spinor basis used, since
it deserves $-$due to its usefulness$-$ to be more widely understood and 
used. Its basic advantage is to extend  smoothly to massive fermions
the natural spin basis for massless fermions.

A well-known covariant way of characterizing spin states is by invoking
the translation $-$invariant Pauli-Lubanski axial 4-vector operator 
\cite{25}.
\begin{eqnarray}
\qquad W^{\mu}=-{1\over 2}\varepsilon^{\mu\nu\lambda\rho}
\Sigma_{\nu\lambda}\widehat{P}_{\rho},
\qquad (\widehat{P}_{\mu}=i\partial_{\mu}) \nonumber \\
W^{\mu}\widehat{P}_{\mu}=0, \qquad [W^{\mu}, \widehat{P}^{\nu}]=0 \\
\qquad [W^{\mu},W^{\nu}]=i\varepsilon^{\mu\nu\lambda\rho}
\widehat{P}_{\lambda} W_{\rho}, \qquad [W^2, W^{\nu}]=0 
\end{eqnarray}
where $\Sigma_{\mu\nu}$ are the spin-parts of Lorentz generators,
$W^2=W_{\mu}W^{\mu}$, and we use the
conventions $\varepsilon^{0123}=1$ and $g_{00}=1$, $g_{11}=g_{22}=
g_{33}=-1$.
$\widehat{P}_{\mu}$ gives $\pm P_{\mu}$ when acting on 
$e^{\mp iP_{\mu}\chi^{\mu}}$, i.e. on particle (antiparticle) solutions
of 4-momentum $P_{\mu}$. For $P^{\mu}$ timelike (massive particle), $W^{\mu}$
is spacelike and reduces in the particle rest-frame to the spin 3-vector
$\vec{\Sigma}$
\begin{eqnarray}
W^{\mu}=(0, \pm m\vec{\Sigma})\quad\hbox{for}\quad P^{\mu}=(m, 0) \nonumber
\end{eqnarray}
while for $P^{\mu}$ lightlike (massless particle) $W^{\mu}$ is lightlike
`parallel' to $P^{\mu}$, 
%and related to the helicity $h=\vec{P}
%\cdot\vec{\Sigma}/P^0$ through 
$W^{\mu}=\pm hP^{\mu}$, where the helicity 
$h={\vec{P}\cdot\vec{W}/(P^0)^2}$ is a Lorentz-invariant
characteristic of a massless particle. According to (3),
(4) we can choose simultaneous eigenstates of $\widehat{P}^{\mu}$,
$W^2$ and of 
the projection of $W^{\mu}$ along some (eventually $P^{\mu}$-dependent)
4-vector $S^{\mu}$: a spin basis is characterized by  the choice of
$S^{\mu}$. This choice is best understood if we complement $P^{\mu}$ by
three independent 4-vectors so as to form a complete basis of Lorentz
4-vectors convenient for expressing $W^{\mu}$.

A $P^{\mu}$ either lightlike or timelike can be complemented by two
unit spacelike 4-vectors $e^{\mu}_{(1)}$, $e^{\mu}_{(2)}$ such that they
form an orthogonal sub-basis ($a\cdot b$ and $a^2$ mean 
$a_{\mu}b^{\mu}$ and $a_{\mu}a^{\mu}$ respectively)
\begin{eqnarray}
P\cdot e_{(1)}=P\cdot e_{(2)}=e_{(1)}\cdot e_{(2)}=0,\qquad
e_{(1)}^2=e_{(2)}^2=-1.
\end{eqnarray}
The choice of the fourth basic 4-vector useful for both $P^2=0$ and 
$P^2>0$ is motivated as follows.

Since for massless particles the helicity $h$ (identical for such particles
to the chirality) is a Poincar\'{e} invariant property conserved by vector
and axial-vector interactions, we choose eigenstates of $h$ as a spinor 
basis for  massless fermions. Thus we choose $S^{\mu}=P^{\mu}$ for
such particles. Complementing now $P^{\mu}$, $e_{(1)}^{\mu}$,
$e_{(2)}^{\mu}$ by a  lightlike 4-vector $k^{\mu}$ such that
\begin{eqnarray}
k\cdot e_{(1)}=k\cdot e_{(2)}=k^2=0,\qquad P\cdot k\not =0
\end{eqnarray}
we can write
\begin{eqnarray}
W^{\mu}=\pm hP^{\mu}={W\cdot k\over P\cdot k}P^{\mu}
\end{eqnarray}
a form convenient for generalization to massive particles.

Turning now to massive particles, $P^{\mu}$ becomes timelike and $W^{\mu}$
spacelike ($W^2=-m^2\Sigma^2$).
 Expressing such a $W^{\mu}$ in our Lorentz basis we find 
(using $W\cdot P=0$ and (5), (6))
\begin{eqnarray}
W^{\mu}={W\cdot k\over P\cdot k}
\left( P^{\mu}-{m^2\over P\cdot k}k^{\mu}\right) -
(W\cdot e_{(1)})e_{(1)}^{\mu}-(W\cdot e_{(2)})e_{(2)}^{\mu}
\end{eqnarray}
Thus choosing for  massive particles
\begin{eqnarray}
S^{\mu}=P^{\mu}-{m^2\over P\cdot k}k^{\mu}
\end{eqnarray}
we obtain a spin-characterization which connects  smoothly the 
massive to the massless case: as $P^2=m^2\to 0$, $W^{\mu}\to \pm hP^{\mu}$,
$S^{\mu}\to P^{\mu}$ and $(8)\to (7)$. This $S^{\mu}$ satisfies
\begin{eqnarray}
S^2=-m^2,\qquad S\cdot P=0.
\end{eqnarray}
If $k^{\mu}$ transforms to $\ell^{\mu}$ when Lorentz-boosting to the rest
frame of the particle, $S^{\mu}$ transforms simultaneously to 
$(0, -m\widehat{\ell})$, where $\widehat{\ell}$ is the unit 3-vector along
$\vec{\ell}$, while $W^{\mu}$ transforms to $(0, \pm m\vec{\Sigma})$.
Hence
\begin{eqnarray}
{W\cdot k\over P\cdot k}=-{W\cdot S\over m^2}=\pm\vec{\Sigma}\cdot
\widehat{\ell}
\end{eqnarray}
i.e. our spinor basis for a massive particle corresponds to eigenstates of 
the projection of the spin $\vec{\Sigma}$ along the unit vector 
$\widehat{\ell}$ in the rest frame of the particle.

The final result of this procedure is to use, for  both massless and 
massive fermions, as  spinor basis the eigenstates of 
$\displaystyle{{W\cdot k\over P\cdot k}}$, where $k^{\mu}$ is some fixed 
lightlike 4-vector such that $P\cdot k\not =0$, and $P^{\mu}$, $W^{\mu}$
are the 4-momentum and the Pauli-Lubanski 4-vector of each particle. It 
is useful to notice that for spin $\displaystyle{{1\over 2}}$ fermions
\begin{eqnarray}
-{1\over 2}\varepsilon^{\mu\nu\lambda\rho}\Sigma_{\nu\lambda}=i\gamma_5
\Sigma^{\mu\rho}
\end{eqnarray}
(with the convention $\gamma_5=i\gamma^0\gamma^1\gamma^2\gamma^3)$,
so that
\begin{eqnarray}
W\cdot k=-{1\over 2}\gamma_5(k\!\!\!/\widehat{P}\!\!\!/
 -\widehat{P}\!\!\!/ k\!\!\!/ )= \gamma_5(\widehat{P}\!\!\!/ k\!\!\!/
 -\widehat{P}\cdot k)
\end{eqnarray}
using $\Sigma^{\mu\nu}={i \over 2}[\gamma^{\mu}, \gamma^{\nu}]$
and the notation $a\!\!\!/ =a_{\mu}\gamma^{\mu}$ for any 4-vector
$a_{\mu}$. In this case our spin-characterizing operator can be written
\begin{eqnarray}
{W\cdot k\over P\cdot k}=\pm
\left({P\!\!\!/ k\!\!\!/\over P\cdot k}-1\right)\gamma_5
\end{eqnarray}
where $+\, (-)$ corresponds to positive (negative) frequency solutions,
i.e. to particle (antiparticle) wavefunctions $u(p)$ $(v(p))$ respectively.

Denoting the eigenvalues of $\displaystyle{{W\cdot k\over P\cdot k}}$
by $\lambda/2$ $(\lambda =\pm 1)$, let $u_{\lambda}(p, m)$,
$v_{\lambda}(p, m)$ be the corresponding fermion and antifermion 
eigenstates, solutions of the Dirac equation with 4-momentum $p_{\mu}$. It
is easy to see that, for  both massive $(m>0)$ and massless
$(m=0)$ particles, all the above requirements and the 
normalization-fixing relations
\begin{eqnarray}
\sum_{\lambda}u_{\lambda}(p, m)\overline{u}_{\lambda}(p, m)=
p\!\!\!/ +m,\qquad
\sum_{\lambda}v_{\lambda}(p, m)\overline{v}_{\lambda}(p, m)=
p\!\!\!/ -m
\end{eqnarray}
are satisfied by the choice \cite{23}
\begin{eqnarray}
u_{\lambda}(p, m)={1\over \eta}(p\!\!\!/ +m)\nu_{-\lambda}(k),\qquad
v_{\lambda}(p, m)=u_{-\lambda}(p, -m)
\end{eqnarray}
where $\eta =\sqrt{2p\cdot k}$ and
\begin{eqnarray}
k\!\!\!/ \nu_{\lambda}(k)& =0,\qquad
\nu_{\lambda}(k)\overline{\nu}_{\lambda}(k)={1\over 2}
(1+\lambda\gamma_5)k\!\!\!/  
\end{eqnarray}
i.e. $\nu_{-}(k)$, $\nu_+(k)$ are the helicity eigenstates of a 
 massless spin $\displaystyle{{1\over 2}}$ fermion having the 
reference 4-momentum $k_{\mu}$. We can take e.g.
$\nu_+(k)=e\!\!\!/ \nu_-(k)$ where $e^{\mu}$ is any 4-vector satisfying
$e^2=-1$, $e\cdot k=0$. One should notice that (17) corresponds to a 
normalization
\begin{eqnarray}
\overline{\nu}_{\lambda}(k)\gamma^{\mu}\nu_{\lambda}(k)=2k^{\mu}
\end{eqnarray}
which, in collaboration with the $1/\eta$ factor in (16), leads to
\begin{eqnarray}
\overline{u}_{\lambda}(p, m) u_{\lambda}(p, m)=2m,\qquad
\overline{v}_{\lambda}(p, m)v_{\lambda}(p, m)=-2m.
\end{eqnarray}
According to our spin specification
\begin{eqnarray}
u_{\lambda}(p, m)\overline{u}_{\lambda}(p, m) ={1\over 2}
[1+\lambda\gamma_5e\!\!\!/_{(3)}](p\!\!\!/ +m) \nonumber \\
={1\over 2}
\left[ 1+\lambda\gamma_5\left( 1-{m\over p\cdot k}k\!\!\!/\right)\right]
(p\!\!\!/ +m)
\end{eqnarray}
where $e_{(3)}^{\mu}$ is the dimensionless unit spacelike 4-vector $S^{\mu}/m$. 
 The corresponding expression for 
$v_{\lambda}(p, m)\overline{v}_{\lambda}(p, m)$ is obtained by changing 
$\lambda\to -\lambda$, $m\to -m$ in the r.h.s. of (20). Formulae (15), (16),
(19) and (20) have  a smooth limit for $m\to 0$, and give then the correct
relations for massless fermions.

Any tree Feynman diagram is characterized by a number of uninterrupted fermion 
lines, each streaming a fermion nb. flow which enters and leaves the diagram.
Along such a line there is a contraction of Dirac indices of a sequence of 
spinors, interaction vertices and numerators of fermion propagators.
Vector or scalar bosom propagators provide contractions of Lorentz
indices of vertices belonging to different fermion lines. 
The essence of the 'spinor bracket' or 'pseudohelicity method' of 
\cite{23},\cite{24} is to express each such fermion line sequence in terms of
few basic blocks.
To exhibit 
the structure of each $iM^F_{\lambda}$ in terms of the building blocks
of the present approach, substitute in the corresponding expression
all numerators 
$p\!\!\!/\pm m$ of fermion propagators according to (15) and 
write all coupling constants as $C_LP_L+C_RP_R$ in terms of 
the chiral projectors
$\displaystyle{P_L={1\over 2}(1-\gamma_5),\,\,\, P_R={1\over 2}
(1+\gamma_5)}.$
Then each $iM^F_{\lambda}$ becomes a sum of
products, these last having as factors two types of complex 
 scalar functions: the spinor brackets
$$\overline{u}_{\lambda_1}(p_1, m_1)(C_LP_L+C_RP_R)
u_{\lambda_2}(p_2, m_2)$$
(whose form motivates the choice of building units, see below), 
and the so called $Z$-functions
\begin{eqnarray}
Z(1, 2, 3, 4) =[\overline{u}_{\lambda_1}(p_1, m_1)
 \Gamma^{\mu}
u_{\lambda_2}(p_2, m_2)][\overline{u}_{\lambda_3}(p_3, m_3)
\Gamma_{\mu}^{\,\prime}
u_{\lambda_4}(p_4, m_4)] \nonumber \\
\Gamma^{\mu} =\gamma^{\mu}(C_LP_L+C_RP_R),\qquad
\Gamma_{\mu}^{\,\prime}=\gamma_{\mu}(C_L'P_L+C_R'P_R) \nonumber
\end{eqnarray}
Decomposing next $u_{\lambda}(p, \pm m)$ into chiral spinors 
$\alpha_{\lambda}(p)$, $\nu_{\lambda}(k)$
$$u_{\lambda}(p, \pm m)={1\over\eta}p\!\!\!/ \nu_{-\lambda}(k)\pm {m\over\eta}
\nu_{-\lambda}(k)\equiv\alpha_{\lambda}(p)\pm\mu\nu_{-\lambda}(k)$$
and using the so-called Chisholm identities for such spinors
\begin{eqnarray}
[\overline{\alpha}_{\lambda}(q_1)\gamma^{\mu}\alpha_{\lambda'}(q_2)]
\gamma_{\mu} =2\delta_{\lambda\lambda'}[\alpha_{\lambda}(q_2)
\overline{\alpha}_{\lambda}(q_1)+\alpha_{-\lambda}(q_1)
\overline{\alpha}_{-\lambda}(q_2)]  \nonumber \\   
\qquad [\overline{\alpha}_{\lambda}(q)\gamma_{\mu}\nu_{\lambda'}(k)]\gamma^{\mu}=
2\delta_{\lambda\lambda'}[\nu_{\lambda}(k)\overline{\alpha}_{\lambda}(q)-
\alpha_{-\lambda}(q)\overline{\nu}_{-\lambda}(k)]  \nonumber \\
\qquad [\overline{\nu}_{\lambda}(k)\gamma_{\mu}\alpha_{\lambda'}(q)]\gamma^{\mu}=
2\delta_{\lambda\lambda'}[\alpha_{\lambda}(q)
\overline{\nu}_{\lambda}(k)-\nu_{-\lambda}(k)
\overline{\alpha}_{-\lambda}(q)] \nonumber
\end{eqnarray}
one can express straightforwardly, and once
for ever \cite{23},\cite{24}, also the $Z$-functions in terms
of the following basic units (here we slightly refine previous work)
\begin{eqnarray}
Y^{L, R}_{\lambda}(p_1, p_2) \equiv\overline{u}_{\lambda}
(p_1)P_{L, R}u_{\lambda}(p_2) \nonumber \\
S^{L, R}_{\lambda}(p_1, p_2) \equiv\overline{u}_{\lambda}
(p_1)P_{L, R}u_{-\lambda}(p_2)=\overline{\alpha}_{\lambda}(p_1)
P_{L, R}\alpha_{-\lambda}(p_2) \nonumber \\ 
\mu={m\over\eta},\qquad \overline{\alpha}_{\lambda}(p)
\nu_{-\lambda}(k)=\overline{\nu}_{\lambda}(k)\alpha_{-\lambda}(p)=
\sqrt{2p\cdot k}\equiv\eta  \nonumber
\end{eqnarray}

The main properties and the explicit expressions (after adopting
$\nu_+(k)=e\!\!\!/ \nu_-(k)$) of the $Y$- and $S$-functions are
\begin{eqnarray}
&Y^L_{\lambda}(p_1, p_2) =Y^R_{-\lambda}(p_1, p_2)=
Y^L_{-\lambda}(p_2, p_1)=Y^R_{\lambda}(p_2, p_1) \nonumber \\
&Y^L_-(p_1, p_2) =\mu_1\eta_2,\quad Y^R_-(p_1, p_2)=\mu_2\eta_1 \nonumber \\
&S^{L, R}_{\lambda}(p_1, p_2) =-S^{L, R}_{\lambda}(p_2, p_1),\quad
S^R_{-\lambda}(p_1, p_2)=-S^L_{\lambda}(p_1, p_2)^{\ast} \nonumber \\
&S^L_+(p_1, p_2) ={2\over \eta_1\eta_2}[(p_1\cdot k)(p_2\cdot e)-
(p_2\cdot k)(p_1\cdot e)+i\varepsilon (k, e, p_1, p_2)] \nonumber \\
&S^R_-(p_1, p_2) =-S^L_+(p_1, p_2)^{\ast},\quad
S^R_+(p_1, p_2)=S^L_-(p_1, p_2)=0 \nonumber
\end{eqnarray}
where $\varepsilon (k, e, p_1, p_2)=\varepsilon_{\mu\nu\lambda\rho}
k^{\mu}e^{\nu}p_1^{\lambda}p_2^{\rho}$, $\varepsilon_{0123}=-1$.
We shall denote the nonzero $S$-functions simply $S_{\pm}(p_1, p_2)$,
neglecting the upper $L, R$ index. For completeness of presentation, the 
expressions for $Z$-functions are tabulated below. For $k^{\mu}$ and 
$e^{\mu}$ we used the convenient choice $k^{\mu}=(1, 1, 0, 0)$,
$e^{\mu}=(0, 0, 1, 0):k^{\mu}$ should be non-orthogonal to all external 
$p^{\mu}$.

\begin{eqnarray}
&Z(+, +, +, +) \  =2[S_+(p_1, p_3)S_-(p_4, p_2)C_RC_R'+
\mu_1\mu_2\eta_3\eta_4 C_LC_R'+\eta_1\eta_2\mu_3\mu_4 C_RC_L'] \nonumber \\
&Z(+, +, -,-) \ =2[S_+(p_1, p_4)S_-(p_3, p_2)C_RC_L'+\mu_1\mu_2\eta_3\eta_4
C_LC_L'+\eta_1\eta_2\mu_3\mu_4 C_RC_R'] \nonumber \\
&Z(+,-, -, +) \ =2[\mu_1\eta_2\mu_3\eta_4 C_LC_R'+\eta_1
\mu_2\eta_3\mu_4C_RC_L'
-\mu_1\eta_2\eta_3\mu_4C_LC_L'-\eta_1\mu_2\mu_3\eta_4C_RC_R'] \nonumber \\
&Z(+,-, +,-)  \ =0 \nonumber \\
&Z(+, +, +,-)  \  =2\eta_2C_R[\mu_3S_+(p_1, p_4)C_L'+\mu_4S_+(p_3, p_1)C_R']
\nonumber \\
&Z(+, +,-, +) \ =2\eta_1C_R[\mu_4S_-(p_3, p_2)C_L'+\mu_3S_-(p_2, p_4)C_R']
\nonumber \\
&Z(+,-, +, +) \ =2\eta_4C_R'[\mu_2S_+(p_1, p_3)C_R+\mu_1S_+(p_3, p_2)C_L]
\nonumber \\
&Z(+,-,-,-) \ =2\eta_3C'_L[\mu_1S_+(p_4, p_2)C_L+\mu_2S_+(p_1, p_4)C_R] 
\nonumber
\end{eqnarray}

The remaining $Z$ are obtained by exchanging simultaneously
$+\leftrightarrow -$ and $L\leftrightarrow R$ in the above expressions.

To recapitulate, using the described spinor basis and following the 
outlined calculational procedure, each Feynman amplitude $iM^F_{\lambda}$
was expressed as a sum of products of the above defined $Y$-, $S$- and 
$Z$-functions. The $Z$-functions are expressed in turn 
 in terms  of the truly basic building units, the $S-$ and  
$Y$-functions, having simple properties and explicit expressions. The whole 
scheme is very convenient and allows compact programming.

We use the t'Hooft-Feynman gauge and
Breit-Wigner  propagators for the Z,W and H bosons.
The mass of W ($m_W = 80.23$ GeV) and the mass ($m_Z=(91.1888$ GeV) and
width ($\Gamma_Z= 2.4974$ GeV) of Z have been taken as inputs;
the width of W being calculated from the SM tree-level formula.

The HB mass was varied in the range 80-120 GeV. For each $m_H$ 
the total $\Gamma_H$ is estimated
using the program kindly provided by the authors of \cite{29}, which
includes the significant $QCD$ and electroweak radiative corrections 
to the tree-level  width formula.
The widths $\Gamma_H$ and $\Gamma_{b\bar{b}}$ found for various
$m_H$ are as follows:

\begin{table}[hbtp]
\begin{center}
\begin{tabular}{|c|c|c|c|c|}
\hline
$m_H$ (GeV)& 80 & 90 & 100 &  110 \\
\hline
$\Gamma_H$ (MeV)
&1.734 & 1.911  & 2.106 & 2.393 \\
\hline
$\Gamma_{b\bar{b}}$ (MeV)
&1.513 & 1.661  & 1.806 & 1.948 \\
\hline
\end{tabular}
\end{center}
\end{table}

In order to include approximatively the corresponding QCD correction to the 
$Hb\bar{b}$ vertex in the Feynman amplitudes, we employ in the  $Hb\bar{b}$
coupling a running mass $\bar{m_b}(m_H^2)$ which, when used in the tree-level
formula for  $\Gamma_{b\bar{b}}$, reproduces the above stated widths. 
In the range of masses considered here, $\bar{m_b}(m_H^2)$ differs from the
QCD running mass at $Q^2=m_H^2$ \cite{27} by a factor 1.12 $\pm$0.01.
The value of $a_S(m_Z^2)$ was taken 0.123. 
The kinematical quark masses needed in various parts of our 
calculations are taken to be $m_b=4.7$, $m_c=1.45$  and $m_t=175$ GeV,
while the KM matrix element $U_{tb}=1$. 
The gauge couplings were fixed by adopting
$\alpha_{eff}=1/128.07$, $(1/137.0359895)$ at $Z(\gamma)$ vertices
and using the effective $g^2=4\sqrt{2} G_F m_W^2$ ($G_F$=1.16639565 10$^{-5}$). 
These effective parameters partly include 
EW radiative corrections in our tree-level amplitudes, in the 
spirit of the `Improved Born Approximation' \cite{28}. 

The QED ISR corrections were calculated approximatively 
using the universal radiation factor, incorporating virtual and soft
contributions up to second order in $\alpha$, with infrared photon
exponentiation, and a hard bremsstrahlung contribution \cite{30}.
Our cross sections should not be sensitive to final state radiation,
because $\nu\bar{\nu}$ do not radiate and $b\bar{b}$ are massive
enough.

Denoting final $\nu, \overline{\nu}$, $b, \overline{b}$ as 
3, 4, 5, 6 respectively, their Lorentz-invariant phase space volume
element $d\hbox{Lips\ }(s; p_3 p_4 p_5 p_6)$ was decomposed 
into a product of effective 2-body elements
\begin{eqnarray}
d\hbox{Lips\ }(s; p_{34} p_{56})d\,\hbox{Lips\ }(s_{34}; p_3 p_4)
d\hbox{Lips\ }(s_{56}; p_5 p_6) {ds_{34}\over 2\pi}\,\,{ds_{56}\over 2\pi}
\nonumber
\end{eqnarray}
where 
$$p^{\mu}_{ij}=p^{\mu}_i+p^{\mu}_j,\qquad s_{ij}=E^2_{ij}-\vec{p}_{ij}^2$$
and  each 2-body $d\,$Lips was evaluated in its c.m. frame. 
Smoothing the integrand by well-known changes of variables \cite{31}, the 
phase space integration was performed by the event generation and Monte
Carlo integration program VEGAS 
\cite{32}, with  estimated accuracy better than 1\%.

The fortran program exists also in a MC generator form, where it is
interfaced to JETSET \cite{jetset} and so after hadronization 
produces final state particles.
\footnote{The code is available from the authors upon request}
The MC generator is controlled by a set of cards where one can 
determine the HB mass, the c.m energy, the inclusion or not of
radiative corrections, possible cuts on $m_{\nu\bar{\nu}}$ and
$m_{b\bar{b}}$, the number of requested events, and whether one wishes to
write the events in LUND format on some separate file.
One can restrict, through switches in the same cards,
the calculation to a subclass of
diagrams and to the production of $\nu_e$ or $\nu_{\mu}$ only. 
Finally, all input constants used in the calculations:  
$m_Z, m_W, \Gamma_Z, G_F$, 
$\alpha$(at 0 and $M_Z$), $\alpha_S$ and the quark kinematical masses,
are given from the outside on the same set of cards,
providing to the user complete control of the input variables for
comparison purposes.

\section{Results and comments}

Considering SM processes $e^+e^-\rightarrow b\bar{b}+f\bar{f}$ at LEPII,
it is well known that about 20\% of the Higgs signal belongs to the
final state $\nu\bar{\nu}b\bar{b}$, imbedded in a large background.
Using the calculational procedure and the parameters presented in the 
previous section, we now proceed to investigate  the characteristic 
properties of the HB-mediated (Fig. 1) and the coherent background
(Fig. 2) contributions to $e^+e^-\to\nu\overline{\nu}b\overline{b}$
(sum over neutrino species), in order to  provide usefull information
for Higgs searches at LEPII.

To gain an orientation we start with Fig. ~\ref{first}, which 
shows, for $m_H=90$ GeV and over c.m. energies
$150<\sqrt{s}\hbox{ (GeV)}<240$, the total expected number of 
$e^+e^-\to\nu\overline{\nu}b\overline{b}$ events 
for integrated luminosity 1000 pb$^{-1}$,  together with the expected 
signal-only events from  only the HB-mediated diagrams of Fig. 1 
and the  expected background events from the diagrams of Fig. 2.
The expected signal events when one includes also the interference term
between diagrams 1 and 2 are also plotted, but are indistinguishable
(differences below 0.2 \%)\footnote{The smallness of this interference
term was checked to hold for all Higgs masses including $m_H=m_Z$} 
from the signal-only curve.
In the same figure the two lower curves provide the signal-to-background ratio
and the signal-to-Higgsstrahlung ratio, the second indicating the 
relative importance of the WW fusion contribution to the signal.
It is interesting to note that the last contribution provides an important
extension of the HB detectability limit at LEPII. We see from the figure
that for $\sqrt{s}$ clearly above the HZ threshold the WW fusion contributes
about 10\% of the signal, but for $\sqrt{s}$ around and below $m_H+m_Z$ 
its relative importance grows rapidly with decreasing c.m energy. We will
examine further down a realistic numerical example.
 
Fig. ~\ref{sig_exc} shows the Higgs 
signal for $m_H$= 80, 90, 100 and 110 GeV and the
background cross sections at LEPII energies, before and after taking into
account ISR corrections. These corrections depend strongly on $\sqrt{s}$, 
ranging from 5\% to  25\% and being largest (as expected) at those 
energies where the cross sections increases most rapidly with $\sqrt{s}$.

We next examine  the relative importance of various 
background contributions and their
energy dependence, paying particular attention to $E_{CM}=175, 192$ 
and $205$ GeV, which are the nominal energies  studied in 
\cite{altarelli}.
Fig. ~\ref{back_exc}  shows the separate contributions 
due to the background processes of Figures 2a, 2b, 2c and 2d, without
and with the ISR corrections respectively. The
$ZZ/\gamma$ cascade contribution (Fig. 2a) dominates. It grows
rapidly at the $ZZ$ production threshold, while the other contributions
increase more gradually with $\sqrt{s}$. 
At $\sqrt{s}=175$, 192 and 205 GeV the total coherent background cross 
section with ISR corrections is about $0.006$ pb, $0.045$ pb and 
$0.065$ pb respectively. It is clear that the diagrams
of Fig. 2d can be practically ignored at LEP II energies, and those 
of Fig. 2b contribute too weakly to provide some sensitivity on the 
exact value of $m_t$. An interesting effect is a very strong destructive
intereference between the WW fusion and W exchange diagrams of the
background, as clearly demonstrated on the same figure. This is a 
reflection of the unitarity (through gauge symmetry) of the SM, which
prevents these background components from violating asymptotically the
$\log(s)$ bound.

To decide about acceptance cuts, it is important to know the more 
accessible differential cross sections of the background and the signal 
contributions. As such we studied the differential cross sections versus
the invisible invariant mass $m_{\nu\bar{\nu}}$,
the visible invariant mass $m_{b\bar{b}}$,
the $\cos\theta_{b\bar{b}}$ of the total
visible 3-momentum $\vec{p}_{b\bar{b}}$ and the  
$\cos\theta_b$ of the $b$-quark 3-momentum
in the collision c.m. system. All differential cross sections have
been calculated with ISR corrections. 
The above differential cross sections for the signal ($m_H$=90 GeV) and the
total background at $\sqrt{s}=192$ GeV are presented on figure 
~\ref{diff_sigs}. The next figure ~\ref{diff_ens} shows how the background 
changes when decreasing $\sqrt{s}$ from 192 to 175 GeV. And Fig. 
~\ref{diff_backs}
shows how the background differential cross sections split into
approximatively gauge independent components ZZ/$\gamma$, WW fusion 
+W exchange and Z radiation.

The gross features of the background can be understood from the 
structure of the $ZZ/\gamma$ cascade diagrams, which dominates it.
The sharp peak around $m_{\nu\nu}\approx m_Z$ is evidently due to all $\nu\overline{\nu}$ pairs
being produced through an intermediate $Z$. The two peaks at 
$m_{b\bar{b}} \geq 0$ 
and $m_{b\bar{b}}\approx m_Z$, are due to the $\gamma\to b\overline{b}$ and 
$Z\to b\overline{b}$ cascades in the diagrams of Fig. 2a. The peak at 
$m_{b\bar{b}}\approx m_Z$ dominates as the energy increases and we pass the
ZZ production threshold. One should also 
notice that the $ZZ/\gamma$ cascade and the $W$ exchange contributions
peak along the beam axis. This is due to the $t$- and $u$-channel
exchange of a light fermion in diagrams of Fig. 2a and 2c, leading in the
differential cross section to factors of the form
$(\sin^2\theta_{bb}+(\hbox{mass})^2/s)^{-1}$,
where $(\hbox{mass})^2/s$ is very small for $Z\gamma$ cascade 
and small for $ZZ$ cascade
(except for $\sqrt{s}$ just above the $ZZ$ production threshold): for 
a discussion see e.g. \cite{33}. 

The peak of the signal around $m_{\nu\bar{\nu}}\approx m_Z$ is due to
the dominance of the Higgsstrahlung diagrams above threshold, while
the WW fusion produces a broad $m_{\nu\bar{\nu}}$ spectrum. The narrow 
width of the HB is responsible for the spectacular peak at $m_{b\bar{b}}$.
One should also notice that, in contrast to  the background,
the signal angular distributions are remarkably isotropic. 
For the Higgsstahlung contribution this is in agreement with the analytic
result \cite{34} that the differential cross section of
$e^+e^-\to HZ$ has angular distribution (in the c.m. system)
$$1+{Q^2\sin^2\theta\over 2m^2_Z}$$
where $Q=\lambda^{1/2}(s, m^2_Z, m^2_H)/2\sqrt{s}$ is the 3-momentum of
$Z$ or $H$: for $\sqrt{s}$ just above $m_H+m_Z$, $Q^2/2m^2_Z$ is very
small and the distribution is flat, but this will change 
sufficiently above the $HZ$ production threshold.
The angular flatness of the $WW$ fusion signal 
is in accordance with the expectation of an isotropic distribution of the 
3-momenta of the Higgs bosons produced by WW fusion at energies not far 
beyond the WW threshold.
It is also clear that there should be considerable constructive interference 
between the two signal components for $\sqrt{s}$ around and below the HZ
threshold. Indeed an examination shows that at these c.m energies about half
the increase of the signal cross section associated with the presence
of WW fusion comes from the interference with the Higgsstrahlung, while at
higher $\sqrt{s}$ this interference dies out.
Similar conclusions have been reached in the very recent study of \cite{zerwas}.

Knowing the main properties of the signal and the coherent background,
one should next proceed to cope with the crucial problem of the very large 
incoherent background, whose main components, as discussed in the introduction,
are $B^{\gamma}_{QCD}$, $B_{WW}$, $B_W$ and $B_{\ell\ell}$
($\ell =e$ and $\mu$). This problem is clearly beyond the scope
of this work: it has been studied by the LEP experimental groups 
as reported in \cite{altarelli}.
We shall simply adopt the following event selection criteria,
which were prescribed by the LEP200 event generator group,
and study how they affect the signal
and the coherent background:

\noindent
(i)\ 50 GeV$< m_{b\bar{b}}$  \quad 
\noindent
(ii)\ ($m_Z$-25 GeV) $ \leq m_{\nu\bar{\nu}} \leq $ ($m_Z$+25 GeV)

Experimental suppression of all background to the HB signal in 
$e^+e^-\to b\overline{b}\,+$ missing products requires of course several
further event acceptance criteria, such as allowing events with only 
two jets, both $b$-tagged and of constrained morphology (acoplanarity and
acollinearity cuts), forbidding evens with too small number of 
charged tracks or with large missing energy in the beam pipe, etc 
\cite{altarelli}. 

The total signal
and background cross sections after cuts (i) and (ii), for various HB masses
and without or with ISR corrections, 
are shown on Figs. ~\ref{sig_exc1}. Fig ~\ref{back_exc1} shows how these
cuts affect the background components.
We see that the coherent background is reduced
greatly (by about 65\%) at $\sqrt{s}=175 GeV$, but the reduction is much 
smaller (10-15\%) at $\sqrt{s}=192$ and 205 GeV. This can be 
attributed mostly to cut (i) which supressed the photon mediated $b\bar{b}$
production, whose relative importance in the total background cross-section
decreases with increasing $\sqrt{s}$. As regards the signal, it is 
mostly influenced by cut (ii), which reduces the contribution induced by
the WW fusion and characterised by a broad $m_{\nu\bar{\nu}}$ spectrum.
So cuts (i)+(ii) lower the signal  by only 
5-10\% for $\sqrt{s} \geq m_H + m_Z$, 
but the reduction grows rapidly 
as $\sqrt{s}$ decreases below the HZ threshold.

It is interesting to estimate carefully, including the effect of our cuts,
how the HB detectability limit at LEPII is extended by the WW  fusion
process. To this end, we show on fig ~\ref{mh100} the $\sqrt{s}$
dependence of the (ISR corrected) signal cross section before and after
cuts, for $m_H=100$ GeV, and the corresponding total signal/Higgsstrahlung 
ratio. We see that for $\sqrt{s}$ clearly above $m_H + m_Z$, the WW
fusion contribution (about 10\% of the signal) is reduced by about 30-40\%
by our cuts. In the region 180-200 GeV, where the WW fusion relative importance
increases rapidly with decreasing $\sqrt{s}$, our cuts 
reduce by only about 20\% the signal-to-Higgsstrahlung ratio. E.g. at
$\sqrt{s}$=192 GeV about 50\% of the events are due to WW fusion, this
reducing to about 40\% after the cuts. The final number of expected events 
for an integrated luminosity (e.g. by summing 4 experiments over 2 years) 
of 1000 pb$^{-1}$ is 8.
 
Finally Fig. ~\ref{btag} presents information which allows an immediate
estimate of the reduction of the number of background and signal events
due to the limited acceptance of micro-vertex detectors used to tag
the b jets. It is clear that signal and coherent background are about
equally lowered by this effect. Micro-vertex detectors extending to
$\cos\theta$=0.9 permit the loss of only about 10-15\% of the events
when b-tagging of both jets is required, while one has a negligible loss
(around 2\%) if contented with only one b-tagged jet.   

In order to allow comparisons with other existing Higgs codes,
we followed the LEP200 Higgs event generator prescriptions 
concerning the input values and obtained the cross-sections 
requested for comparison purposes. We present them in the tables below.
As can be seen in \cite{altarelli} 
the agreement is satisfactory, differences are of the order of  1\%.

\begin{table}[hbtp]
\begin{center}
\begin{tabular}{|c|c|c|c|c|}
\hline
$m_H$~(GeV)& 65 & 90 & 115 &  $\infty$ \\
\hline
175 GeV ($\nu_{\mu}{\bar\nu_{\mu}}$) &63.447(0.289)& 2.316(0.011)&1.257(0.006) &1.257(0.006) \\
\hline
192 GeV ($\nu_{\mu}{\bar\nu_{\mu}}$) &72.01(0.317)&46.840(0.093)&19.575(0.087)&19.453(0.087) \\
\hline
205 GeV ($\nu_{\mu}{\bar\nu_{\mu}}$) &67.704(0.301)&53.569(0.240) &25.531(0.114)& 23.646(0.105) \\
\hline
175 GeV ($\nu_e{\bar \nu_e}$)        &70.103(0.302)&4.996(0.023)&1.051(0.006)&1.051(0.006) \\
\hline
192 GeV ($\nu_e{\bar \nu_e}$) &79.158(0.348)&52.744(0.140)&20.605(0.102)&19.678(0.098) \\
\hline
205 GeV ($\nu_e{\bar \nu_e}$) &76.165(0.359)&61.429(0.280)&31.048(0.146)&25.737(0.122) \\
\hline
\end{tabular}
\end{center}
\caption{The process $ e^+ e^- \to \nu {\bar \nu} b \bar b $ at Born level,
following the LEP200 generator Higgs group input values and prescriptions.
Cross sections in fb.The numerical integration errors are included in 
parentheses.}
\end{table}

\begin{table}[hbtp]
\begin{center}
\begin{tabular}{|c|c|c|c|c|}
\hline
$m_H$~(GeV)& 65 & 90 & 115 &  $\infty$ \\
\hline
175 GeV ($\nu_{\mu}{\bar\nu_{\mu}}$) 
&54.186(0.283)& 1.660(0.008)&0.917(0.006) &0.917(0.006) \\
\hline
192 GeV ($\nu_{\mu}{\bar\nu_{\mu}}$) 
&64.405(0.246)&35.367(0.109)&14.68(0.087)&14.60(0.087) \\
\hline
205 GeV ($\nu_{\mu}{\bar\nu_{\mu}}$) 
&63.826(0.222)&45.736(0.171) &21.275(0.105)& 20.058(0.105) \\
\hline
175 GeV ($\nu_e{\bar \nu_e}$) 
&60.806(0.294)&3.663(0.018) &0.775(0.006)  &0.775(0.006) \\
\hline
192 GeV ($\nu_e{\bar \nu_e}$) 
&71.151(0.268)&40.212(0.164)&15.222(0.098)&14.61(0.098) \\
\hline
205 GeV ($\nu_e{\bar \nu_e}$) 
&71.794(0.266)&52.543(0.199)&25.428(0.124)&21.76(0.122) \\
\hline
\end{tabular}
\end{center}
\caption{The process $ e^+ e^- \to \nu {\bar \nu} b \bar b $ including ISR
corrections, following the LEP200 generator 
Higgs group input values and prescriptions.
Cross sections in fb.
The numerical integration errors are included in parentheses.}
\end{table}

\clearpage

\section{Summary and conclusions}
In this work we studied the SM process 
$e^+e^- \rightarrow \nu\bar{\nu}b\bar{b}$ (sum over neutrino species) at c.m
energies $\sqrt{s}$= 150-240 GeV and for Higgs boson masses $m_H$=80-120 GeV.
The results were obtained from all tree level diagrams and included 
approximatively the most important radiative corrections, i.e the QED ISR 
corrections, the QCD and EW corrections to the Higgs width and a QCD 
correction to the H$b\bar{b}$ coupling. The matrix elements were calculated
by the 'spinor bracket' method without neglecting masses, and we tried to 
make the method more accessible by describing it in detail and motivating
its spinor basis in a novel way. The phase space integrals were calculated
by an importance sampling Monte Carlo numerical integrator. 

The signal is composed of the Higgsstrahlung and the WW fusion amplitudes.
The first dominates for $\sqrt{s}$ above the HZ threshold, but the second
becomes increasingly important as $\sqrt{s}$ decreases around and below that
threshold, extending probably (depending on luminosity) to 
$ m_H \geq \sqrt{s}-m_Z$ the Higgs detectability limit at LEPII. The two
amplitudes behave similarly over phase space, providing in particular
isotropic angular distributions of visible momenta, but differ in their
$m_{\nu\bar{\nu}}$ spectrum at higher $\sqrt{s}$:
the one of the WW fusion is always broad, while that of the Higgsstrahlung
peaks strongly around $m_{\nu\bar{\nu}}=m_Z$ as soon as 
$\sqrt{s} \geq m_H + m_Z$. In fact the two amplitudes interfere 
semiconstructively for $\sqrt{s}$ below and around 
the HZ threshold, but hardly so well above it. 
About 50\% of the signal cross section around that threshold
is associated with the WW fusion, this becoming about 10\% at higher 
energies. 

The background consists of 4 classes of amplitudes, describing ZZ/$\gamma$
cascade, WW fusion, W exchange and Z radiation processes. The ZZ and 
Z$\gamma$ cascades dominate, the first leading beyond he ZZ threshold.
Between the WW fusion and W exchange components there is a strong 
destructive interference and their coherent contribution is more than
an order of magnitude  smaller than the ZZ/$\gamma$ one, while the Z radiation
contribution is practically negligible.
The overall behaviour over phase space is characterised by a peak around
$m_{\nu\bar{\nu}}=m_Z$, two peaks in $m_{b\bar{b}}$ due to the
$\gamma \rightarrow b\bar{b}$ and $Z \rightarrow b\bar{b}$ cascades,
and a clear angular peak along the beam axis.
The total background cross section (before cuts) lies between
the $m_H=90$ GeV and $m_H=100$ GeV signal cross sections throughout
LEPII energies.
The signal-background interference is negligible, even around $m_H=m_Z$.

Applying a set of minimal cuts prescribed by the LEPII groups, 
$m_{b\bar{b}} \geq 50$ GeV and $m_Z-25$ GeV $ \leq m_{\nu\bar{\nu}} \leq $
$m_Z+25$ GeV, has a very different effect for $\sqrt{s}$ below and 
above the HZ threshold. At lower energies both the signal and the 
background cross sections are greatly reduced, although the first
mostly due to the $m_{\nu\bar{\nu}}$ cut while the second to the
$m_{b\bar{b}}$ cut. At higher energies both the signal and the background
are reduced weakly ($\sim$ 10 \%), the signal-background ratio 
improving only slightly.The effect of an angular acceptance of a
b-tagging micro-vertex on the signal and the background were found
roughly equal, rejecting about 10-15\% of events for present type
extended microvertices.
The code exists also in the form of a MC generator.

\begin{figure}[htb]
\begin{center}
%\mbox{\epsfxsize=15.0cm,\epsfysize=10.0cm\epsffile{fig1.ps}}
\caption{Signal Feynman diagrams}
\label{fig1}
\end{center}
\end{figure}

\begin{figure}[htb]
\begin{center}
%\mbox{\epsfxsize=15.0cm,\epsfysize=15.0cm\epsffile{fig2.ps}}
\caption{Background Feynman diagrams}
\label{fig2}
\end{center}
\end{figure}

\begin{figure}[htb]
\begin{center}
%\mbox{\epsfxsize=15.0cm\epsffile{first.ps}}
\caption{Number of expected $\nu\bar{\nu}b\bar{b}$ events from 
the background, the Higgs signal ($m_H$= 90 GeV) and their sum,
as a function of E$_{CM}$ for 1000 pb$^{-1}$. 
The signal+interference is also plotted, but is 
indistinguishable from the signal-only curve. 
The ratios of the signal-to-background
and the total signal-to-Higgstrahlung contribution are also
shown. QED ISR corrections are included}
\label{first}
\end{center}
\end{figure}

\begin{figure}[htb]
\begin{center}
%\mbox{\epsfxsize=15.0cm\epsffile{sig_exc.ps}}
\caption{Higgs signal $\nu\bar{\nu}b\bar{b}$  
cross section as a function of E$_{CM}$
for  different HB masses and 
the background, without and with ISR corrections}
\label{sig_exc}
\end{center}
\end{figure}

\begin{figure}[htb]
\begin{center}
%\mbox{\epsfxsize=15.0cm\epsffile{backg_exc.ps}}
\caption{Total background $\nu\bar{\nu}b\bar{b}$ cross section
and subprocess contributions 
as functions of E$_{CM}$ without and with ISR corrections}
\label{back_exc}
\end{center}
\end{figure}

\begin{figure}[htb]
\begin{center}
%\mbox{\epsfxsize=15.0cm\epsffile{diff_sigs.ps}}
\caption{Differential cross sections for the Higgs signal with $m_H$ 90 GeV 
(solid line) and for the background (dashed line) 
at $E_{CM}$=192 GeV.(ISR corrections included)}
\label{diff_sigs}
\end{center}
\end{figure}

\begin{figure}[htb]
\begin{center}
%\mbox{\epsfxsize=15.0cm\epsffile{diff_ens.ps}}
\caption{Differential cross sections for the background at 192 GeV 
(solid line) and 175 GeV (dashed line) (ISR corrections included)}
\label{diff_ens}
\end{center}
\end{figure}

\begin{figure}[htb]
\begin{center}
%\mbox{\epsfxsize=15.0cm\epsffile{diff_backs.ps}}
\caption{Differential cross sections, at $E_{CM}$=192 GeV for the total background 
(solid line) and the subprocesses a) ZZ/$\gamma$ production (dashed line)
b) W fusion and exchange (dotted dashed line) 
and c) Z radiation (dotted line) (ISR corrections included)}
\label{diff_backs}
\end{center}
\end{figure}

\begin{figure}[htb]
\begin{center}
%\mbox{\epsfxsize=15.0cm\epsffile{backg_exc1.ps}}
\caption{Total background  $\nu\bar{\nu}b\bar{b}$ cross section
and subprocess contributions
as functions of E$_{CM}$, without and with ISR corrections, after cuts}
\label{back_exc1}
\end{center}
\end{figure}

\begin{figure}[htb]
\begin{center}
%\mbox{\epsfxsize=15.0cm\epsffile{sig_exc1.ps}}
\caption{Higgs signal $\nu\bar{\nu}b\bar{b}$ cross section as a function 
of E$_{CM}$ for different Higgs masses and
the background, without and with ISR corrections, after cuts}
\label{sig_exc1}
\end{center}
\end{figure}

\begin{figure}[htb]
\begin{center}
%\mbox{\epsfxsize=15.0cm\epsffile{mh100.ps}}
\caption{The number of expected Higgs $\nu\bar{\nu}b\bar{b}$
events for 1000 pb$^{-1}$  as a function of E$_{CM}$
for $m_H$ =100 GeV, before and after cuts.
The ratios of the signal-to-background
and the total signal-to-Higgstrahlung contribution are also
shown (ISR corrections are included)}
\label{mh100}
\end{center}
\end{figure}

\begin{figure}[htb]
\begin{center}
%\mbox{\epsfxsize=15.0cm\epsffile{btag.ps}}
\caption{Percentage of events for Higgs (dashed and dotted dashed lines) 
and background (solid and dotted) outside the fiducial acceptance of 
a microvertex extending to cos$\theta$ when 
demanding a) at least one taged jet or b) both jets tagged}
\label{btag}
\end{center}
\end{figure}


\begin{thebibliography}{99}

\bibitem{1}
P.S. Wells, Proceedings of the Brussels Conference, August 1995.

\bibitem{2}CDF Collaboration, F. Abe et al., Phys. Rev. Lett. 73 (1994)
225, Phys. Rev. Lett. 74 (1995) 2626, D0 Collaboration S. Abachi et al.
Phys. Rev. Lett. 74(1995) 2422,2632.


\bibitem{4}K. Hagiwara, D. Haidt, C.S. Kim and S. Matsumoto,
 Z. Phys. C64 (1994) 559, and references therein.
W. Hollik in: Precision Tests of the Standard Model, P.
Langacker editor (Advanced Series on Directions in High Energy Physics,
World Scientific, 1993).

\bibitem{5}G. Altarelli, CERN preprint CERN-TH 7464/94 (1994) and
references therein.


\bibitem{6}Contribution of H.E. Haber in: Perspectives on 
Higgs physics, G.L. Kane editor (World Scientific, 1993).
An extensive previous review, with outdated by now data, is:
J.F. Gunion, H.E. Haber, G. Kane and S. Dawson: The Higgs Hunter's Guide
(Frontiers in Physics, Addison-Wesley, 1990).

\bibitem{6a}Contribution of M. Sher in: Perspectives on 
Higgs physics, G.L. Kane editor (World Scientific, 1993).


\bibitem{7}G. Altarelli and G. Isidori, Phys. Lett. B337 (1994) 141

\bibitem{8}J. Ellis, G.L. Fogli and E. Lisi, CERN preprint CERN-TH
7261/94 (1994).
J.A. Casas, J.R. Espinosa, M. Quiros and A. Riotto,
NUcl. Phys. B436 (1995) 3

\bibitem{9}T. Elliot, S.F. King and P.L. White, Univ. of Southampton
preprint SHEP 92/93-11.

\bibitem{10}M. Carena and C.E.M. Wagner, CERN preprint, CERN-TH 
7393/94 (1994).


\bibitem{14}J. Boucrot et al., presented by Sau Lan Wu: Search for Neutral
Higgs at LEP 200, Proceedings of the ECFA Workshop on LEP 200, Aachen,
1986, published as CERN 87-08, ECFA 87/100 (1987).

\bibitem{15}D. Bardin et al., convenor S. Katsanevas: Report from the Working
Group on LEP 200 Physics, DELPHI 92-166  PHYS 250, December 1992.

\bibitem{16}J. Alcaraz, M. Felcini, M. Pieri, B. Zhou, CERN preprint
CERN-PPE/93-28, February 1993.

\bibitem{17}D.R.T. Jones and S.T. Petcov, Phys. Lett. B84 (1979) 108.

\bibitem{18}R. Bates and J.N. Ng, Phys. Rev. D32 (1985) 51.

\bibitem{19}D.A. Dicus and S.S.D. Willebrock, Phys. Rev. D32 (1985) 1642.

\bibitem{20}S. Dawson and J.L. Rosner, Phys. Lett. B148 (1984) 497.

\bibitem{21}K. Hikasa, Phys. Lett. B164 (1985) 385.

\bibitem{22a}E. Boos, M. Sachwitz, H.J. Schreiber and S. Shichanin,
DESY preprint 93-183 (1993).

\bibitem{22b}M. Dubinin, V. Edneral, Y. Kurihara and Y. Shimizu, Phys.
Lett. B329 (1994) 379.

\bibitem{23}F.A. Berends, P.H. Daverveldt and R. Kleiss, Nucl. Phys.
B253 (1985) 441.
R. Kleiss and W.J. Stirling, Nucl. Phys. B266 (1985) 235.
P.H. Daverveldt, Ph. D. thesis, Leiden Univ. (1985).

\bibitem{24}C. Mana and M. Martinez, Nucl. Phys. B287 (1987) 601.

\bibitem{25}J.K. Lubanski, Physica 9 (1942) 310.

\bibitem{29}E. Gross, B.A. Kniehl and G. Wolf, Z. Phys. C63 (1994) 417.

\bibitem{27}N. Gray, D.J. Broadhurst, W. Grafe and K. Schilcher, Z. 
Phys. C48 (1990) 673,\\
B.A. Kniehl, Phys. Rep. 240 (1994) 211, 

\bibitem{28}M. Consoli, W. Hollik, F. Jegerlehner in: Z Physics at LEP I,
G. Altarelli, R. Kleiss, C. Verzegnassi (editors), CERN Yellow Report
No. 89-08 (1989), Vol. 1, p. 7.\\
A. Denner in: Proceedings of the 28$^{th}$ Rencontres de Moriond:
Electroweak Interactions and Unified Theories, Editor J.Tran Thanh Van,
Editions Frontieres, 1994, p. 433.

\bibitem{30} Program REMT by R. Kleiss.

\bibitem{31}E.g. V.D. Barger and R.J.N. Phillips: Collider Physics 
(Frontiers in Physics Series, Addison-Wesley, 1987), Chapter 11.

\bibitem{32} G.P. Lepage J. Comput. Phys. 27  (1978) 192. 

\bibitem{33}F.M. Renard: Basics of electron positron collisions 
(Editions Fronti\`{e}res, 1981), Chapter 4.

\bibitem{34}R.L. Kelly and T. Shimada, Phys. Rev. D23 (1981) 1940.

\bibitem{zerwas} W.Kilian, M. Kramer and P.M.Zerwas DESY 95-216, hep-ph/9512355.

\bibitem{altarelli} CERN Yellow report CERN-TH/95-151, CERN-PPE/95-78
G. Altarelli, T.Sjostrand, F. Zwirner. Eds. 

\bibitem{jetset} T. Sjostrand, Comput. Phys. Commun. 39 (1986) 347;
T. Sjostrand and M. Bengstsson, Comput. Phys. Commun. 43 (1987) 367;
 H.U.Bengstsson and T. Sjostrand,  Comput. Phys. Commun. 46 (1987) 43.

\end{thebibliography}
\end{document}